\documentclass[aps,prb,twocolumn]{revtex4-2}

\usepackage{bm,color}
\usepackage[dvipdfmx]{graphicx}
\usepackage{amsmath, amssymb}
\usepackage{tabularx}
\usepackage{braket}
\begin{document}
\title{Multiple Floquet Chern insulator phases in the spin-charge coupled triangular-lattice ferrimagnet: Crucial role of higher-order terms in the high-frequency expansion}
\author{Rintaro Eto}
\affiliation{Department of Applied Physics, Waseda University, Okubo, Shinjuku-ku, Tokyo 169-8555, Japan}
\author{Masahito Mochizuki}
\affiliation{Department of Applied Physics, Waseda University, Okubo, Shinjuku-ku, Tokyo 169-8555, Japan}
\date{\today} 
\begin{abstract}
We study the effects of photoirradiation with circularly polarized light on the Dirac half-metal state induced by the ferrimagnetic order in a triangular Kondo-lattice model. Our analysis based on the Floquet theory reveals that two types of Floquet Chern insulator phases appear as photoinduced nonequilibrium steady states and that these two phases can be experimentally detected and distinguished by measurements of the Hall conductivity. It is elucidated that these rich nonequilibrium topological phases come from higher-order terms in the high-frequency expansion called Brillouin-Wigner expansion, which is in striking contrast to usually discussed Floquet Chern insulator phases originating from the lowest-order terms of the expansion. So far, the lattice electron models on simple non-multipartite lattices such as triangular lattices and square lattices have not been regarded as targets of the Floquet engineering because the lowest-order terms of the high-frequency expansion for Floquet effective Hamiltonians cancel each other to vanish in these systems. Our findings of the Floquet Chern insulator phases in a triangular Kondo-lattice model are expected to expand the range of potential models and even materials targeted by the Floquet engineering.
\end{abstract}
\maketitle
\section{Introduction}
Since the first theoretical proposal of the Floquet Chern insulator state in graphene~\cite{Oka2009,Kitagawa2011}, the Floquet engineering of topology in condensed matters has been a central issue of the photoinduced physics~\cite{Oka2019,Torre2021}. 
%%The predicted photoinduced Hall current of topological origin in graphene was indeed observed in experiments recently~\cite{McIver2020,SatoSA2019}. 
A recent experiment~\cite{McIver2020} and a precise theoretical analysis~\cite{SatoSA2019} indicate that, although the experimentally observed photoinduced transverse current in graphene is largely coming from the electron population imbalance around the Dirac cone, there is tiny but finite contribution from photoinduced Berry curvatures.
Possible realizations of the Floquet Chern insulator state are theoretically and experimentally proposed also in several atomic-layer systems including twisted multilayer graphene~\cite{Topp2019,Rodriguez-Vega2020,Assi2021}, silicene~\cite{Ezawa2013}, black phosphorene~\cite{Kang2020}, transition-metal dichalcogenides~\cite{Claassen2016,ZhangMY2019,Chono2020,Vogl2021}, organic salts~\cite{Kitayama2020,Kitayama2021a,Kitayama2021b,Kitayama2022,Kitayama2023,Tanaka2021,Tanaka2022}, and cuprate superconductors~\cite{Takasan2017}. In addition, subsequent seminal works have revealed new concepts, e.g., the anomalous Floquet states due to hybridization of multiple Floquet sectors with different photon numbers~\cite{Rudner2013,Perez-Piskunow2014,Usaj2014,Roy2017} and the Floquet fractional Chern insulator states as intrinsic fractionally quantized Hall states under a periodic drive~\cite{Grushin2014}, which are attracting a great deal of research interest recently. The circular dichroism of the Floquet states is also studied intensively~\cite{Schuler2020,Schuler2022}. 

The Floquet engineering of material topology has been developed further in various directions~\cite{Rudner2020a}. On the surfaces of topological insulators, Floquet-Bloch states with ultrafast intraband and interband dynamics are experimentally observed~\cite{WangYH2013,Ito2023}. In three-dimensions, Floquet Weyl semimetal states~\cite{WangR2014,Ebihara2016} as three-dimensional counterparts of the Floquet Chern insulator states have been proposed to emerge in the photodriven three-dimensional Dirac electron systems~\cite{Hubener2017} and the Mott-insulating magnets on the pyrochlore lattice~\cite{Topp2018}. Consideration of strong electron correlations is another important direction of the research. The Floquet theory combined with the dynamical mean-field theory (DMFT) named Floquet-DMFT has been developed as a powerful tool to analyze the photodriven phenomena in correlated electron systems~\cite{Tsuji2008,Aoki2014}, e.g., manipulation of fermion-fermion interactions~\cite{Tsuji2011} and higher-harmonic generations in Mott insulators~\cite{Murakami2018}. In addition, the Floquet engineering extends its scope to the research field of spintronics~\cite{Takayoshi2014,Tanaka2020}. Analyses based on the Floquet theory have revealed a variety of physical mechanisms of spin-current generation based on spin-wave excitations by application of electromagnetic waves to magnets such as multiferroic materials with magnetoelectric coupling~\cite{Sato2016}. Spin dynamics toward and in the Floquet states are also active research topics~\cite{Ono2017,Ono2018}. The Floquet engineering continues to attract researchers from the viewpoints of both fundamental science and technical application.

The geometry of lattice structure is a crucial factor for the Floquet engineering~\cite{Kitagawa2011,Mikami2016,Ikeda2023}. We argue this aspect by introducing a technical detail of the Floquet theory. The Floquet theory describes nonequilibrium steady states induced by a periodic drive such as light in an extended Hilbert space of states with different photon numbers $n$, i.e., the original space without the periodic drive ($n$=0) and duplicated subspaces with $n=\pm1, \pm2, \cdots$. Because direct analyses of this extended Hilbert space require a huge computational cost, a high-frequency expansion technique is frequently used, in which the duplicated nonzero photon-number subspaces are projected onto the original Hilbert space. In this framework, we often consider the terms up to the first-order of the expansion with respect to $1/\omega$ for the effective Hamiltonian $\hat{H}_{\rm eff}$, 
%%%%%%%%%%%%%%%%%%%%%%%%%%%%%%%%%%%%%%%
\begin{align}
&\hat{H}_{\rm eff}=\hat{H}_0 - \sum_{n>0} \frac{\left[ \hat{H}_n,\hat{H}_{-n} \right]}{n\omega}, 
\label{eq:Heff} 
\\
&\hat{H}_n=\frac{1}{T}\int_0^T d\tau \ \hat{H}(\tau)e^{in\omega\tau},
\label{eq:Heff2}
\end{align}
%%%%%%%%%%%%%%%%%%%%%%%%%%%%%%%%%%%%%%%
where $\hat{H}(\tau)$ denotes the original time-dependent Hamiltonian under photoirradiation, and $\omega$ and $T=2\pi/\omega$ denote the frequency and time period of light, respectively. The first term of Eq.~(\ref{eq:Heff}) describes the time-average of the original Hamiltonian, while the second term describes the projection of one-photon subspaces into the original Hilbert space. 

In fact, in the Floquet engineering, the second term plays a substantial role for the emergence of photoinduced topological phases. However, it is known that the first-order terms usually vanish because of the cancellation of contributions from equivalent paths having phases with opposite signs. In fact, we can avoid such cancellations to realize a nonzero contribution from the first-order terms in multipartite lattices, e.g, honeycomb lattices (bipartite), Lieb lattices (bipartite), and Kagome lattices (tripartite). On the contrary, the cancellation cannot be avoided in simple non-multipartite lattices such as square and triangular lattices even in the presence of extrinsic sublattice degrees of freedom introduced by long-range orders of spins and/or charges. For this reason, lattice electron models on several multipartite lattices such as Kagome lattices, honeycomb lattices, and Lieb lattices have been intensively studied in the research of Floquet engineering, but those on simple square lattices and triangular lattices have not been its major target so far.

Then we encounter the following question: Is it really impossible to conduct the Floquet engineering with electron models on simple non-multipartite lattices? The answer is no, and the Floquet engineering can indeed be applied to lattice electron models on, e.g., triangular and square lattices. In this paper, we demonstrate that even a simple triangular-lattice system can host rich Floquet topological phases under photoirradiation based on a theoretical study on the photodriven Dirac half-metal state in a Kondo-lattice model with spin-charge coupling. Our Floquet analysis reveals that originally massless half-metallic Dirac electrons become massive under irradiation with circularly polarized light, and, consequently, two kinds of Floquet spin-polarized Chern insulator phases emerge. Importantly, the first-order terms of high-frequency expansion are not relevant to this photoinduced topological phase transition because they completely cancel each other out and vanish on the triangular lattice. Instead, higher-order terms turn out to play a substantial role for emergence of these Floquet topological phases. We also argue that these Floquet Chern insulator phases can be detected and distinguished by a measurement of the Hall conductivity. Our findings are expected to broaden the range of candidate materials for the Floquet engineering.

The rest of this paper is organized as follows. In Sec.~II, we describe the extended Ising Kondo-lattice model utilized in this study. In Sec.~III, we explain our theoretical framework based on the Floquet theory and the high-frequency expansion. In this section, we also discuss how the first-order terms of the high-frequency expansion vanish in the non-multipartite lattice systems. In Sec.~IV, we present the results, which include nonequilibrium band structures, nonequilibrium phase diagrams, and the Hall conductivity under irradiation with circularly polarized light. Section~V is devoted to summary.

\section{Model}
%%%%%%%%%%%%%%%%%%%%%%%%%%%%%%%%%%%%%%%
\begin{figure}[t]
\centering
\includegraphics[scale=0.5]{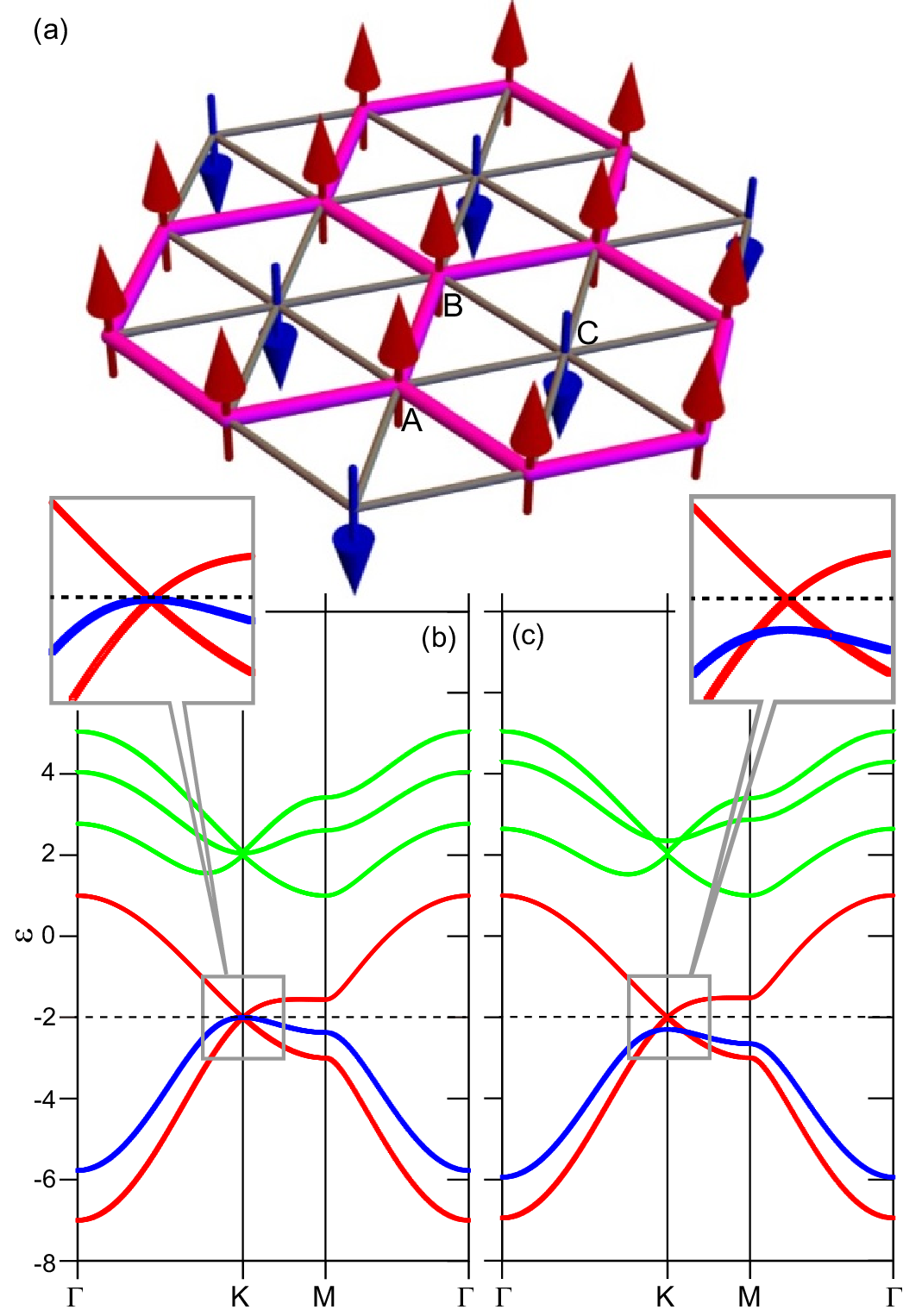}
\caption{(a) Up-up-down ferrimagnetic order with three sublattices (A, B, and C) on the triangular lattice. Thick lines indicate a honeycomb network of up-spins. (b) Band dispersion relations for the extended Ising Kondo-lattice model with $J_{\rm K}/t=2$ and $J'_{\rm K}/t=0$. (c) Those for $J_{\rm K}/t=2$ and $J'_{\rm K}/t=0.05$. Red and blue colors of the lowest three bands in (b) and (c) indicate the up- and down-spin polarizations, respectively.}
\label{Fig01}
\end{figure}
%%%%%%%%%%%%%%%%%%%%%%%%%%%%%%%%%%%%%%%
We start with a triangular Kondo-lattice model which describes the coupling between conduction electrons and localized spins on the triangular lattice~\cite{Ishizuka2012}. The Hamiltonian is given by
%%%%%%%%%%%%%%%%%%%%%%%%%%%%%%%%%%%%%%%
\begin{align}
\mathcal{H}_{\rm IKLM}
&=-t\sum_{\braket{i,j}}\sum_{\sigma=\uparrow,\downarrow} \left(\hat{c}^\dagger_{i\sigma}\hat{c}_{j\sigma} + {\rm h.c.} \right) 
\notag \\
&-J_{\rm K} \sum_{i}\sum_{\sigma,\sigma'=\uparrow,\downarrow} \hat{c}^\dagger_{i\sigma}\hat{\sigma}^z_{\sigma\sigma'}\hat{c}_{i\sigma'}S_{iz}
\notag \\
&+J'_{\rm K} \sum_{i}\sum_{j\in\left\{ {\rm NN} \ {\rm of} \ i \right\}}\sum_{\sigma,\sigma'=\uparrow,\downarrow} \hat{c}^\dagger_{i\sigma}\hat{\sigma}^z_{\sigma\sigma'}\hat{c}_{i\sigma'}S_{jz}.
\label{eq:IKLM}
\end{align}
%%%%%%%%%%%%%%%%%%%%%%%%%%%%%%%%%%%%%%%
Here $\hat{c}^\dagger_{i\sigma}$ $(\hat{c}_{i\sigma})$ denotes the creation (annihilation) operator of an conduction electron with spin $\sigma(=\uparrow,\downarrow)$ on site $i$, and $\bm S_i=(0, 0, S_{iz})$ with $S_{iz}=\pm 1$ is the localized Ising spin on site $i$. The symbol $\hat{\sigma}^z$ is the Pauli matrix for the spin $z$-component. The first term represents the nearest-neighbor hoppings of conduction electrons, where $t$ is the transfer integral and is used as a unit of energy. The second and third terms describe the intrasite ferromagnetic and intersite antiferromagnetic exchange coupling between the conduction electron spins and the localized Ising spins, respectively, where $J_{\rm K}$ and $J'_{\rm K}$ are the coupling constants. We assume a three-sublattice up-up-down ferrimagnetic order of the localized spins [Fig.~\ref{Fig01}(a)]. This spin structure is known to be realized as the ground state when the electron filling is nearly 1/3. We assume the limit of strong Ising anisotropy where this magnetic order does not alter under photoirradiation.

In equilibrium, this ferrimagnetic order induces Dirac-cone bands when $(J_{\rm K}+3J'_{\rm K})/t>1$, whose Dirac points are located at K-point and $\varepsilon=\pm J_{\rm K}$. These Dirac-cone bands are perfectly spin polarized. Specifically, the lower-lying Dirac cone around $\varepsilon=-J_{\rm K}$ is up-spin polarized, whereas the higher-lying Dirac cone around $\varepsilon=+J_{\rm K}$ is down-spin polarized [Fig.~\ref{Fig01}(b)]. When the electron filling is $n_e=1/3$, the lowest two of the six bands are occupied and the lower Dirac point with up-spin polarization is located at the Fermi level. This state is refered to as ``Dirac half metal''. 

In fact, an additional down-spin band overlaps this Dirac point, which inevitably hinders physical responses of the Dirac electrons. The intersite antiferromagnetic Kondo coupling $J'_{\rm K}$ can remove this down-spin band from the Dirac point through lowering its energy as shown in Fig.~\ref{Fig01}(c)~\cite{Ishizuka2012}. This band shift does not change the chemical potential $\mu=-J_{\rm K}$, which is originally located at the Dirac point of lower-lying Dirac-cone bands. In this situation, we expect physical responses purely from the Dirac electrons, at least, at low temperatures. In the following, we discuss the results obtained for $J_{\rm K}/t=2$ and $J_{\rm K}'/t=0.05$. Note that this parameter set does satisfy the relation $(J_{\rm K}+3J'_{\rm K})/t>1$.

When the order of localized spins $\{\bm S_i\}$ is rigid and is never affected by the photoirradiation, the above Kondo-lattice model can be reduced to a tight-binding model with sublattice-dependent on-site potentials as,
%%%%%%%%%%%%%%%%%%%%%%%%%%%%%%%%%%%%%%%
\begin{equation}
\mathcal{H}= -t \sum_{i\neq j,\sigma}
\hat{c}^\dagger_{i\sigma}\hat{c}_{j\sigma} 
+ \sum_{i,\sigma} v_{i,\sigma} \hat{c}^\dagger_{i\sigma}\hat{c}_{i\sigma},
\label{eq:ModelHm}
\end{equation}
%%%%%%%%%%%%%%%%%%%%%%%%%%%%%%%%%%%%%%%
with
%%%%%%%%%%%%%%%%%%%%%%%%%%%%%%%%%%%
\begin{align}
v_{i,\sigma}=
\begin{cases}
-J_{\rm K}\sigma & i \in {\rm sublattice\,A}\\
-J_{\rm K}\sigma & i \in {\rm sublattice\,B}\\
(J_{\rm K}+6J'_{\rm K})\sigma & i \in {\rm sublattice\,C}
\end{cases}
\end{align}
%%%%%%%%%%%%%%%%%%%%%%%%%%%%%%%%%%%
where $\sigma=+1$ ($\sigma=-1$) for up (down) spin. This formula is a general form of the Hamiltonian for the noninteracting electron systems with arbitrary on-site potentials.

When this system is irradiated by light, the situation is described by the following Hamiltonain with a time-dependent vector potential ${\bm A}(\tau)$,
%%%%%%%%%%%%%%%%%%%%%%%%%%%%%%%%%%%%%%%
\begin{equation}
\hat{H}(\tau)= -t \sum_{i\neq j,\sigma}e^{i{\bm A}(\tau)\cdot{\bm r}_{ij}}
\hat{c}^\dagger_{i\sigma}\hat{c}_{j\sigma} 
+ \sum_{i,\sigma} v_{i,\sigma} \hat{c}^\dagger_{i\sigma}\hat{c}_{i\sigma},
\label{eq:FloquetModelHm}
\end{equation}
%%%%%%%%%%%%%%%%%%%%%%%%%%%%%%%%%%%%%%%
where $\bm r_{ij} \equiv \bm r_j-\bm r_i$ denotes the bond vector from site $i$ to site $j$. The general form of the time-dependent vector potential is given by,
%%%%%%%%%%%%%%%%%%%%%%%%%%%%%%%%%%%%%%%
\begin{align}
\bm A(\tau)=\frac{E_\omega}{\omega}\left(\bm e_1 \cos\omega\tau + \bm e_2 \sin\omega\tau \right),
\end{align}
%%%%%%%%%%%%%%%%%%%%%%%%%%%%%%%%%%%%%%%
where $\bm e_1$ and $\bm e_2$ are arbitrary three-component vectors describing the light polarization. This vector potential generates the light electric field,
%%%%%%%%%%%%%%%%%%%%%%%%%%%%%%%%%%%%%%%
\begin{align}
\bm E(\tau)=-\frac{\partial {\bm A}(\tau)}{\partial\tau} 
=E_\omega \left(\bm e_1 \sin\omega\tau - \bm e_2 \cos\omega\tau \right).
\end{align}
%%%%%%%%%%%%%%%%%%%%%%%%%%%%%%%%%%%%%%%
In this study, we examine the effect of circularly polarized light and thus set $\bm e_1=(1,0,0)$ and $\bm e_2=(0,1,0)$.

\section{Methods}
We analyze the model in Eq.~(\ref{eq:FloquetModelHm}) using the Floquet theory. We also perform real-time simulations based on the time-dependent Schr\"{o}dinger equation to support the Floquet analysis. In the following, we describe a fundamental formalism of the Floquet theory and its application to noninteracting systems. Then we describe the Keldysh Green's functions to calculate the Hall conductivity in the photoirradiated system in Sec.~III.B. In Sec.~III.C, we describe the Brillouin-Wigner expansion as one of the typical high-frequency expansion techniques. In Sec.~III.D, we discuss how the first-order terms of the Brillouin-Wigner expansion are canceled in the non-multipartite lattice electron systems. In Sec.~III.E, we explain details of the real-time simulations. We also discuss the unit conversions used in this study in Sec.~III.F.

\subsection{Floquet theory}
The time-dependent Schr\"{o}dinger equation for noninteracting systems is given by
%%%%%%%%%%%%%%%%%%%%%%%%%%%%%%%%%%%%%%%
\begin{equation}
i\hbar\frac{\partial}{\partial \tau}\ket{\psi(\tau)} = \hat{H}(\tau)\ket{\psi(\tau)},
\label{eq:Schrodinger}
\end{equation}
%%%%%%%%%%%%%%%%%%%%%%%%%%%%%%%%%%%%%%%
where $\tau$, $\hat{H}(\tau)$ and $\ket{\psi(\tau)}$ denote the time, the noninteracting time-dependent Hamiltonian, and the time-dependent single-particle wavefunction, respectively. When we consider a time-periodic Hamiltonian $\hat{H}(\tau)$ with a frequency $\omega$, which satisfies $\hat{H}(\tau)=\hat{H}(\tau+T)$ with $T=2\pi/\omega$, the wavefunction $\ket{\psi(\tau)}$ is given by,
%%%%%%%%%%%%%%%%%%%%%%%%%%%%%%%%%%%%%%%
\begin{equation}
\ket{\psi(\tau)} = e^{-i\varepsilon t/\hbar}\ket{\phi(\tau)}.
\label{eq:FloquetState}
\end{equation}
%%%%%%%%%%%%%%%%%%%%%%%%%%%%%%%%%%%%%%%
Here $\ket{\phi(\tau)}(=\ket{\phi(\tau+T)})$ is the time-periodic single-particle state called Floquet state, which has the same time periodicity as that of $\hat{H}(\tau)$, and $\varepsilon$ denotes the eigenvalue of the corresponding Floquet state. Equation~(\ref{eq:FloquetState}) is the representative formula of the Floquet theorem. Because both $\hat{H}(\tau)$ and $\ket{\phi(\tau)}$ are time periodic, we can expand them by the complex Fourier series as,
%%%%%%%%%%%%%%%%%%%%%%%%%%%%%%%%%%%%%%%
\begin{align}
&\hat{H}(\tau)=\sum_m e^{-im\omega\tau} \hat{H}_m, 
\label{eq:complexFourierH} \\
&\ket{\phi(\tau)}=\sum_m e^{-im\omega\tau} \ket{\phi^m}, 
\label{eq:complexFourierWF}
\end{align}
%%%%%%%%%%%%%%%%%%%%%%%%%%%%%%%%%%%%%%%
where the complex Fourier coefficients $\hat{H}_m$ and $\ket{\phi^m}$ are, respectively, given by,
%%%%%%%%%%%%%%%%%%%%%%%%%%%%%%%%%%%%%%%
\begin{align}
&\hat{H}_m=\frac{1}{T}\int_0^T d\tau \ \hat{H}(\tau) e^{im\omega\tau}, 
\\
&\ket{\phi^m}=\frac{1}{T}\int_0^T d\tau \ \ket{\phi(\tau)} e^{im\omega\tau}.
\end{align}
%%%%%%%%%%%%%%%%%%%%%%%%%%%%%%%%%%%%%%%%
Here the integer $m$ in $\hat{H}_m$ and $\ket{\phi^m}$ denotes the number of photons. 

\begin{widetext}
For $\hat{H}(\tau)$ in Eq.~(\ref{eq:FloquetModelHm}), we obtain an explicit formula of the Fourier coefficient $\hat{H}_n$ as,
%%%%%%%%%%%%%%%%%%%%%%%%%%%%%%%%%%%%%%%
%%\begin{align}
%%\begin{aligned}
%%\hat{H}_n &= -t \sum_{i\neq j,\sigma} \mathcal{J}_{-n}(\mathcal{A}_{ij}) e^{-in\theta_{ij}} \hat{c}^\dagger_{i\sigma}\hat{c}_{j\sigma} \\
%%&\quad\quad + \delta_{0,n} \sum_{i,\sigma} v_{i\sigma} \hat{c}^\dagger_{i\sigma}\hat{c}_{i\sigma}.
%%\label{eq:FCHm}
%%\end{aligned}
%%\end{align}
\begin{align}
\hat{H}_n = -t \sum_{i\neq j,\sigma} \mathcal{J}_{-n}(\mathcal{A}_{ij}) e^{-in\theta_{ij}} \hat{c}^\dagger_{i\sigma}\hat{c}_{j\sigma}
+ \delta_{0,n} \sum_{i,\sigma} v_{i\sigma} \hat{c}^\dagger_{i\sigma}\hat{c}_{i\sigma}.
\label{eq:FCHm}
\end{align}
%%%%%%%%%%%%%%%%%%%%%%%%%%%%%%%%%%%%%%%
We use the following relation in the derivation,
%%%%%%%%%%%%%%%%%%%%%%%%%%%%%%%%%%%%%%%
%%\begin{align}
%%&\frac{1}{T} \int_0^T d\tau \exp\left\{
%%i\mathcal{A}_{ij}\sin(\omega \tau+\theta_{ij})+in\omega\tau \right\}
%%\notag \\
%%&=\mathcal{J}_{-n}(\mathcal{A}_{ij}) e^{-in\theta_{ij}},
%%\end{align}
\begin{align}
\frac{1}{T} \int_0^T d\tau \exp\left\{
i\mathcal{A}_{ij}\sin(\omega \tau+\theta_{ij})+in\omega\tau \right\}
=\mathcal{J}_{-n}(\mathcal{A}_{ij}) e^{-in\theta_{ij}},
\end{align}
%%%%%%%%%%%%%%%%%%%%%%%%%%%%%%%%%%%%%%%
where $\mathcal{J}_n(x)$ is the $n$-th order Bessel function of the first kind. The quantities $\mathcal{A}_{ij}$ and $\theta_{ij}$ are, respectively, given by,
%%%%%%%%%%%%%%%%%%%%%%%%%%%%%%%%%%%%%%%
\begin{align}
&\mathcal{A}_{ij}=\frac{E_\omega}{\omega} 
\sqrt{(\bm r_{ij} \cdot \bm e_1)^2 + (\bm r_{ij} \cdot \bm e_2)^2}, 
\\
&\theta_{ij}=\tan^{-1} \left( \frac{\bm r_{ij} \cdot \bm e_1}{\bm r_{ij} \cdot \bm e_2} \right).
\end{align}
%%%%%%%%%%%%%%%%%%%%%%%%%%%%%%%%%%%%%%%

Substituting Eqs.~(\ref{eq:FloquetState}), (\ref{eq:complexFourierH}), and (\ref{eq:complexFourierWF}) into Eq.~(\ref{eq:Schrodinger}), we obtain the following eigenvalue equation,
%%%%%%%%%%%%%%%%%%%%%%%%%%%%%%%%%%%%%%%
\begin{equation}
\left(\hat{\mathcal{H}}_{\rm F} - \omega\hat{\mathcal{M}_{\rm r}}\right) \ket{\phi_{\rm F}}
=\varepsilon \ket{\phi_{\rm F}},
\label{eq:FloquetSambe}
\end{equation}
%%%%%%%%%%%%%%%%%%%%%%%%%%%%%%%%%%%%%%%
where
%%%%%%%%%%%%%%%%%%%%%%%%%%%%%%%%%%%%%%%
\begin{equation} \def\udots{\mathinner{\mkern1mu\raise1pt\vbox{\kern7pt\hbox{.}}\mkern2mu\raise4pt\hbox{.}\mkern2mu\raise7pt\hbox{.}\mkern1mu}}
\hat{\mathcal{H}}_{\rm F} = \left(
\begin{array}{ccccccc}
\ddots & \vdots & \vdots & \vdots & \vdots & \vdots & \udots \\
\cdots & \hat{H}_{-2,-2}
& \hat{H}_{-2,-1}
& \hat{H}_{-2, 0}
& \hat{H}_{-2,+1}
& \hat{H}_{-2,+2} & \cdots \\
\cdots & \hat{H}_{-1,-2}
& \hat{H}_{-1,-1}
& \hat{H}_{-1, 0}
& \hat{H}_{-1,+1}
& \hat{H}_{-1,+2} & \cdots \\
\cdots & \hat{H}_{ 0,-2}
& \hat{H}_{ 0,-1}
& \hat{H}_{ 0, 0}
& \hat{H}_{ 0,+1}
& \hat{H}_{ 0,+2} & \cdots \\
\cdots & \hat{H}_{+1,-2}
& \hat{H}_{+1,-1}
& \hat{H}_{+1, 0}
& \hat{H}_{+1,+1}
& \hat{H}_{+1,+2} & \cdots \\
\cdots & \hat{H}_{+2,-2}
& \hat{H}_{+2,-1}
& \hat{H}_{+2, 0}
& \hat{H}_{+2,+1}
& \hat{H}_{+2,+2} & \cdots \\
\udots & \vdots & \vdots & \vdots & \vdots & \vdots & \ddots
\end{array}
\right),
\end{equation}
%%%%%%%%%%%%%%%%%%%%%%%%%%%%%%%%%%%%%%%
%%%%%%%%%%%%%%%%%%%%%%%%%%%%%%%%%%%%%%%
\begin{equation}
\def\udots{\mathinner{\mkern1mu\raise1pt\vbox{\kern7pt\hbox{.}}\mkern2mu\raise4pt\hbox{.}\mkern2mu\raise7pt\hbox{.}\mkern1mu}}
\hat{\mathcal{M}_{\rm r}} = \left(
\begin{array}{ccccccc}
\ddots & \vdots & \vdots & \vdots & \vdots & \vdots & \udots \\
\cdots & -2\hat{I}_{\rm r}
& \hat{O}_{\rm r}
& \hat{O}_{\rm r}
& \hat{O}_{\rm r}
& \hat{O}_{\rm r} & \cdots \\
\cdots & \hat{O}_{\rm r}
& -\hat{I}_{\rm r}
& \hat{O}_{\rm r}
& \hat{O}_{\rm r}
& \hat{O}_{\rm r} & \cdots \\
\cdots & \hat{O}_{\rm r}
& \hat{O}_{\rm r}
& \hat{O}_{\rm r}
& \hat{O}_{\rm r}
& \hat{O}_{\rm r} & \cdots \\
\cdots & \hat{O}_{\rm r}
& \hat{O}_{\rm r}
& \hat{O}_{\rm r}
& +\hat{I}_{\rm r}
& \hat{O}_{\rm r} & \cdots \\
\cdots & \hat{O}_{\rm r}
& \hat{O}_{\rm r}
& \hat{O}_{\rm r}
& \hat{O}_{\rm r}
& +2\hat{I}_{\rm r} & \cdots \\
\udots & \vdots & \vdots & \vdots & \vdots & \vdots & \ddots \\
\end{array}
\right).
\end{equation}
%%%%%%%%%%%%%%%%%%%%%%%%%%%%%%%%%%%%%%%
In the real-space representation, the matrix $\mathcal{H}_{\rm F}$ is composed of $2N \times 2N$-dimensional block matrices $\hat{H}_{n,m} \equiv \hat{H}_{n-m}$. The matrices $\hat{I}_{\rm r}$ and $\hat{O}_{\rm r}$ are $2N \times 2N$-dimensional identity matrix and zero matrix, respectively.
The Floquet-state vector $\ket{\phi_{\rm F}}$ in the real-space representation is defined as,
%%%%%%%%%%%%%%%%%%%%%%%%%%%%%%%%%%%%%%%
\begin{equation}
\ket{\phi_{\rm F}}=
^{\rm t}(\cdots, \{\phi^{-2}_{i\sigma}\}, \{\phi^{-1}_{i\sigma}\}, \{\phi^{0}_{i\sigma}\}, 
\{\phi^{1}_{i\sigma}\}, \{\phi^{2}_{i\sigma}\}, \cdots),
\end{equation}
%%%%%%%%%%%%%%%%%%%%%%%%%%%%%%%%%%%%%%%
where $\{\phi^n_{i\sigma}\}$ is a set of $2N$ components $\phi^n_{i\sigma}$ of the $n$-photon Floquet-state vector $\ket{\phi^n}$ with $i=1,2,\cdots,N$ and $\sigma=\pm1$. In this way, the time-dependent Schr\"{o}dinger equation in Eq.~(\ref{eq:Schrodinger}) with a time-periodic Hamiltonian $\hat{H}(\tau)$ is mapped onto the time-independent eigenvalue problem given by Eq.~(\ref{eq:FloquetSambe}). 

Using the Fourier transforms,
%%%%%%%%%%%%%%%%%%%%%%%%%%%%%%%%%%%%%%%
\begin{align}
\hat{c}_{i\sigma}^\dagger =\frac{1}{\sqrt{N}}\sum_{\bm k} \hat{c}_{\bm k}^\dagger e^{-i\bm k \cdot \bm r_i},
\quad\quad
\hat{c}_{i\sigma}=\frac{1}{\sqrt{N}}\sum_{\bm k} \hat{c}_{\bm k} e^{i\bm k \cdot \bm r_i},
\end{align}
%%%%%%%%%%%%%%%%%%%%%%%%%%%%%%%%%%%%%%%
we rewrite Eq.~(\ref{eq:FloquetSambe}) in the momentum-space representation as,
%%%%%%%%%%%%%%%%%%%%%%%%%%%%%%%%%%%%%%%
\begin{equation}
\left(\hat{\mathcal{H}}_{\rm F}(\bm k) - \omega\hat{\mathcal{M}_{\rm m}}\right) 
\ket{\phi_{\rm F}(\bm k)}
=\varepsilon_{\bm k} \ket{\phi_{\rm F}(\bm k)},
\label{eq:FloquetSambe2}
\end{equation}
%%%%%%%%%%%%%%%%%%%%%%%%%%%%%%%%%%%%%%%
where
%%%%%%%%%%%%%%%%%%%%%%%%%%%%%%%%%%%%%%%
\begin{equation}
\def\udots{\mathinner{\mkern1mu\raise1pt\vbox{\kern7pt\hbox{.}}\mkern2mu\raise4pt\hbox{.}\mkern2mu\raise7pt\hbox{.}\mkern1mu}}
\hat{\mathcal{H}}_{\rm F}(\bm k) = \left(
\begin{array}{ccccccc}
\ddots & \vdots & \vdots & \vdots & \vdots & \vdots & \udots \\
\cdots & \hat{H}_{-2,-2}(\bm k)
& \hat{H}_{-2,-1}(\bm k)
& \hat{H}_{-2, 0}(\bm k)
& \hat{H}_{-2,+1}(\bm k)
& \hat{H}_{-2,+2}(\bm k) & \cdots \\
\cdots & \hat{H}_{-1,-2}(\bm k)
& \hat{H}_{-1,-1}(\bm k)
& \hat{H}_{-1, 0}(\bm k)
& \hat{H}_{-1,+1}(\bm k)
& \hat{H}_{-1,+2}(\bm k) & \cdots \\
\cdots & \hat{H}_{ 0,-2}(\bm k)
& \hat{H}_{ 0,-1}(\bm k)
& \hat{H}_{ 0, 0}(\bm k)
& \hat{H}_{ 0,+1}(\bm k)
& \hat{H}_{ 0,+2}(\bm k) & \cdots \\
\cdots & \hat{H}_{+1,-2}(\bm k)
& \hat{H}_{+1,-1}(\bm k)
& \hat{H}_{+1, 0}(\bm k)
& \hat{H}_{+1,+1}(\bm k)
& \hat{H}_{+1,+2}(\bm k) & \cdots \\
\cdots & \hat{H}_{+2,-2}(\bm k)
& \hat{H}_{+2,-1}(\bm k)
& \hat{H}_{+2, 0}(\bm k)
& \hat{H}_{+2,+1}(\bm k)
& \hat{H}_{+2,+2}(\bm k) & \cdots \\
\udots & \vdots & \vdots & \vdots & \vdots & \vdots & \ddots
\end{array}
\right),
\end{equation}
%%%%%%%%%%%%%%%%%%%%%%%%%%%%%%%%%%%%%%%
%%%%%%%%%%%%%%%%%%%%%%%%%%%%%%%%%%%%%%%
\begin{equation}
\def\udots{\mathinner{\mkern1mu\raise1pt\vbox{\kern7pt\hbox{.}}\mkern2mu\raise4pt\hbox{.}\mkern2mu\raise7pt\hbox{.}\mkern1mu}}
\hat{\mathcal{M}_{\rm m}} = \left(
\begin{array}{ccccccc}
\ddots & \vdots & \vdots & \vdots & \vdots & \vdots & \udots \\
\cdots & -2\hat{I}_{\rm m}
& \hat{O}_{\rm m}
& \hat{O}_{\rm m}
& \hat{O}_{\rm m}
& \hat{O}_{\rm m} & \cdots \\
\cdots & \hat{O}_{\rm m}
& -\hat{I}_{\rm m}
& \hat{O}_{\rm m}
& \hat{O}_{\rm m}
& \hat{O}_{\rm m} & \cdots \\
\cdots & \hat{O}_{\rm m}
& \hat{O}_{\rm m}
& \hat{O}_{\rm m}
& \hat{O}_{\rm m}
& \hat{O}_{\rm m} & \cdots \\
\cdots & \hat{O}_{\rm m}
& \hat{O}_{\rm m}
& \hat{O}_{\rm m}
& +\hat{I}_{\rm m}
& \hat{O}_{\rm m} & \cdots \\
\cdots & \hat{O}_{\rm m}
& \hat{O}_{\rm m}
& \hat{O}_{\rm m}
& \hat{O}_{\rm m}
& +2\hat{I}_{\rm m} & \cdots \\
\udots & \vdots & \vdots & \vdots & \vdots & \vdots & \ddots \\
\end{array}
\right).
\end{equation}
%%%%%%%%%%%%%%%%%%%%%%%%%%%%%%%%%%%%%%%
The matrix $\mathcal{H}_{\rm F}(\bm k)$ is composed of $6 \times 6$-dimensional block matrices $\hat{H}_{n,m}(\bm k) \equiv \hat{H}_{n-m}(\bm k)$. The matrices $\hat{I}_{\rm m}$ and $\hat{O}_{\rm m}$ are $6 \times 6$-dimensional identity matrix and zero matrix, respectively.
The Floquet-state vector $\ket{\phi_{\rm F}(\bm k)}$ in the momentum space is defined as,
%%%%%%%%%%%%%%%%%%%%%%%%%%%%%%%%%%%%%%%
\begin{equation}
\ket{\phi_{\rm F}(\bm k)}=
^{\rm t}(\cdots, \{\phi^{-2}_\nu(\bm k)\}, \{\phi^{-1}_\nu(\bm k)\}, \{\phi^{0}_\nu(\bm k)\}, 
\{\phi^{1}_\nu(\bm k)\}, \{\phi^{2}_\nu(\bm k)\}, \cdots),
\end{equation}
%%%%%%%%%%%%%%%%%%%%%%%%%%%%%%%%%%%%%%%
where $\nu$ is the band index. The present system has six bands for each photon-number subspace (i.e., $\nu=1,2,\cdots,6$) because of the three sublattices and the spins $\sigma=\pm1$. Here $\{\phi^n_\nu(\bm k)\}$ is a set of six components $\phi^n_\nu(\bm k)$ of the $n$-photon Floquet-state vector $\ket{\phi^n(\bm k)}$ at momentum $\bm k$. For practical treatment of Eq.~(\ref{eq:FloquetSambe2}), we restrict the number of photons to $|n| \le n_{\rm max}$ with $n_{\rm max}$=16 throughout the present study. After this truncation, we obtain the band dispersion relations $\varepsilon_\nu^n(\bm k)$ and the eigenstates $\ket{\phi_\nu^n(\bm k)}$ by diagonalizaing the truncated Floquet Hamiltonian $\mathcal{H}_{\rm F}(\bm k)-\omega\mathcal{M}_{\rm m}$.
\end{widetext}

\subsection{Keldysh Green's function formalism}
The Chern number of the $\nu$th band in the Floquet state and the Hall conductivity under photoirradiation are respectively given by~\cite{Oka2009,Rudner2020a,Rudner2020b},
%%%%%%%%%%%%%%%%%%%%%%%%%%%%%%%%%%%%%%%
\begin{align}
&N_{\rm Ch}^\nu=\frac{1}{2\pi} \int_{\rm BZ} d{\bm k} \ \Omega_z^{\nu}({\bm k}), \\
&\sigma_{xy}=\frac{e^2}{h} \frac{1}{2\pi} \sum_\nu \int_{\rm BZ} d{\bm k} \ n_{\nu}(\bm k) \Omega_z^{\nu}(\bm k).
\end{align}
%%%%%%%%%%%%%%%%%%%%%%%%%%%%%%%%%%%%%%%
The Berry curvature $\Omega_z^{\nu}({\bm k})$ is given by,
%%%%%%%%%%%%%%%%%%%%%%%%%%%%%%%%%%%%%%%
\begin{widetext}
\begin{align}
\Omega_z^{\nu}(\bm k) =i \sum_{(m,\mu)\left[\neq(n,\nu)\right]} \frac{\braket{\phi_\nu^n(\bm k)|\partial_{k_x}\mathcal{H}_{\rm F}(\bm k)|\phi_\mu^m(\bm k)} \braket{\phi_\mu^m(\bm k)|\partial_{k_y}\mathcal{H}_{\rm F}(\bm k)|\phi_\nu^n(\bm k)}}{\left[ \varepsilon_\nu^n(\bm k) - \varepsilon_\mu^m(\bm k) \right]^2}.
\end{align}
\end{widetext}
%%%%%%%%%%%%%%%%%%%%%%%%%%%%%%%%%%%%%%%
The nonequilibrium electron distribution function $n_\nu(\bm k)$ represents the expectation value of electron occupation of the state with a momentum $\bm k$ and the band index $\nu$. For the off-resonant condition that all the Floquet sidebands with nonzero photon number $n\neq 0$ do not overlap the original band set with $n=0$, we can simply approximate $n_\nu(\bm k)\sim f(\varepsilon_\nu^0(\bm k))$ where $f(\varepsilon)$ is the Fermi distribution function in equilibrium. On the contrary, for the on-resonant condition that some of the Floquet sidebands overlap the original bands, this approximation no longer holds. Therefore, we need to calculate $n_\nu(\bm k)$ for evaluating the 
Hall conductivity.

We utilize Keldysh Green's function method~\cite{Tsuji2009,Aoki2014} to calculate $n_\nu({\bm k})$. The Dyson equation for the Green's function matrix is given by,
%%%%%%%%%%%%%%%%%%%%%%%%%%%%%%%%%%%%%%%
\begin{align}
\begin{aligned}
&\quad \left(
\begin{array}{cc}
\hat{G}^{\rm R}({\bm k},\varepsilon) & \hat{G}^{\rm K}({\bm k},\varepsilon) \\
0 & \hat{G}^{\rm A}({\bm k},\varepsilon)
\end{array}
\right)^{-1} \\ 
&= \left(
\begin{array}{cc}
\left[ \hat{G}^{\rm R0}({\bm k},\varepsilon) \right]^{-1} & \left[ \hat{G}^{\rm K0}({\bm k},\varepsilon) \right]^{-1} \\
0 & \left[ \hat{G}^{\rm A0}({\bm k},\varepsilon) \right]^{-1}
\end{array}
\right) - \left(
\begin{array}{cc}
\hat{\Sigma}^{\rm R} & \hat{\Sigma}^{\rm K}(\varepsilon) \\
0 & \hat{\Sigma}^{\rm A}
\end{array}
\right).
\end{aligned}
\end{align}
%%%%%%%%%%%%%%%%%%%%%%%%%%%%%%%%%%%%%%%
where $\hat{G}^{\rm R}$, $\hat{G}^{\rm A}$, and $\hat{G}^{\rm K}$ ($\hat{\Sigma}^{\rm R}$, $\hat{\Sigma}^{\rm A}$, and $\hat{\Sigma}^{\rm K}$) are the retarded, advanced, and Keldysh Green's functions (self-energies) for the Floquet states, respectively. Matrix elements of the noninteracting Green's functions for the Floquet states and the self-energies are respectively given by,
%%%%%%%%%%%%%%%%%%%%%%%%%%%%%%%%%%%%%%%
\begin{align}
&[\hat{G}^{\rm R0}({\bm k},\varepsilon)]^{-1}_{n\nu,m\mu}=\varepsilon\delta_{nm}\delta_{\nu\mu} - [\mathcal{H}_{\rm F}({\bm k})-\omega\mathcal{M}_{\rm m}]_{n\nu,m\mu},
\\
&[\hat{G}^{\rm R0}({\bm k},\varepsilon)]^{-1}_{n\nu,m\mu}=\varepsilon\delta_{nm}\delta_{\nu\mu} - [\mathcal{H}_{\rm F}({\bm k})-\omega\mathcal{M}_{\rm m}]_{n\nu,m\mu},
\\
&[\hat{\Sigma}^{\rm R}]_{n\nu,m\mu}=-i\Gamma\delta_{nm}\delta_{\nu\mu},
\\
&[\hat{\Sigma}^{\rm A}]_{n\nu,m\mu}= i\Gamma\delta_{nm}\delta_{\nu\mu},
\\
&[\hat{\Sigma}^{\rm K}(\varepsilon)]_{n\nu,m\mu}= -2i\Gamma \tanh\left[ \frac{\varepsilon-\mu+m\omega}{2k_{\rm B}T_{\rm hr}} \right] \delta_{nm}\delta_{\nu\mu},
\end{align}
%%%%%%%%%%%%%%%%%%%%%%%%%%%%%%%%%%%%%%%
where $\Gamma$ represents the strength of dissipation due to the coupling to a heat reservoir at temperature $T_{\rm hr}$. We set $\Gamma/t=0.1$ in this study. Then, the lesser Green's function $\hat{G}^<$ is given by,
%%%%%%%%%%%%%%%%%%%%%%%%%%%%%%%%%%%%%%%
\begin{equation}
\hat{G}^<({\bm k},\varepsilon)=
\hat{G}^{\rm R}({\bm k},\varepsilon)\hat{\Sigma}^<(\varepsilon)\hat{G}^{\rm A}({\bm k},\varepsilon),
\end{equation}
%%%%%%%%%%%%%%%%%%%%%%%%%%%%%%%%%%%%%%%
where the lesser self-energy $\hat{\Sigma}^<$ is given by,
%%%%%%%%%%%%%%%%%%%%%%%%%%%%%%%%%%%%%%%
\begin{equation}
\hat{\Sigma}^<(\varepsilon)=\frac{\hat{\Sigma}^{\rm A} + \hat{\Sigma}^{\rm K}(\varepsilon) - \hat{\Sigma}^{\rm R}}{2}.
\end{equation}
%%%%%%%%%%%%%%%%%%%%%%%%%%%%%%%%%%%%%%%
Finally, the nonequilibrium electron distribution $n_\nu(\bm k)$ with momentum $\bm k$ and band index $\nu$ is given by,
%%%%%%%%%%%%%%%%%%%%%%%%%%%%%%%%%%%%%%%
\begin{equation}
 n_\nu(\bm k) =
\frac{\braket{\phi^0_\nu(\bm k)|\hat{N}_{\bm k}\left(\varepsilon^0_\nu(\bm k)\right)|\phi^0_\nu(\bm k)}}
       {\braket{\phi^0_\nu(\bm k)|\hat{A}_{\bm k}\left(\varepsilon^0_\nu(\bm k)\right)|\phi^0_\nu(\bm k)}},
\end{equation}
%%%%%%%%%%%%%%%%%%%%%%%%%%%%%%%%%%%%%%%
where the operators $\hat{A}$ and $\hat{N}$ are respectively given by,
%%%%%%%%%%%%%%%%%%%%%%%%%%%%%%%%%%%%%%%
\begin{align}
&\hat{A}_{\bm k}(\varepsilon)= \frac{i}{2\pi} \left( \hat{G}^{\rm R}({\bm k},\varepsilon) - \hat{G}^{\rm A}({\bm k},\varepsilon) \right), 
\\
&\hat{N}_{\bm k}(\varepsilon)= -\frac{i}{2\pi} \hat{G}^<({\bm k},\varepsilon).
\end{align}
%%%%%%%%%%%%%%%%%%%%%%%%%%%%%%%%%%%%%%%
In this study, we use $144\times144$ ${\bm k}$-points for numerical calculations. The chemical potential $\mu$ is determined by using the bisection method to preserve the total electron number $n_{\rm e}$. We iterate the Keldysh Green's function calculation to tune the chemical potential until the value of $n_{\rm e}$ becomes very close to the target value. In the following, we keep $n_{\rm e}=0.34$ with negligibly small errors less than $10^{-8}$.

\subsection{Brillouin-Wigner expansion}
To study the effect of lattice geometry on the electron states in the photoirradiated system, we adopt the Brillouin-Wigner expansion~\cite{Mikami2016}, which is one of the typical high-frequency expansion techniques~\cite{Rahav2003,Hausinger2010,Goldman2014,Eckardt2015}. The high-frequency expansion is accurate in the limit of $\omega\rightarrow\infty$ where $2\pi/\omega$ is a time periodicity of the Hamiltonian. The expansion gives an effective Hamiltonian $\hat{H}_{\rm BW}$ called Brillouin-Wigner Hamiltonian, which describes the electron states under photoirradiation and is obtained by an appropriate projection of nonzero photon-number subspaces of $\mathcal{H}_{\rm F}-\omega\mathcal{M}_{\rm r}$ to the original Hilbert space with zero photon number. The Brillouin-Wigner Hamiltonian is given by,
%%%%%%%%%%%%%%%%%%%%%%%%%%%%%%%%%%%%%%%
\begin{align}
\hat{H}_{\rm BW}=\sum_{n=0,1,2,\cdots} \hat{H}_{\rm BW}^{(n)},
\label{eq:BWE} 
\end{align}
%%%%%%%%%%%%%%%%%%%%%%%%%%%%%%%%%%%%%%%
with
%%%%%%%%%%%%%%%%%%%%%%%%%%%%%%%%%%%%%%%
\begin{align}
&\hat{H}_{\rm BW}^{(0)}=\hat{H}_{0,0}, 
\label{eq:BW0}
\\
&\hat{H}_{\rm BW}^{(1)}=\sum_{n\neq0} \frac{\hat{H}_{0,n}\hat{H}_{n,0}}{n\omega} = -\sum_{n>0} \frac{\left[ \hat{H}_n,\hat{H}_{-n} \right]}{n\omega},
\label{eq:BW1}
\\
&\hat{H}_{\rm BW}^{(2)}=\sum_{n_1,n_2\neq0} \frac{\hat{H}_{0,n_1}\hat{H}_{n_1,n_2}\hat{H}_{n_2,0}}{n_1n_2\omega^2} - \sum_{n\neq0} \frac{\hat{H}_{0,n}\hat{H}_{n,0}\hat{H}_{0,0}}{n^2\omega^2}, 
\label{eq:BW2}
\\
&\hat{H}_{\rm BW}^{(3)}=\sum_{n_1,n_2,n_3\neq 0} \frac{\hat{H}_{0,n_1}\hat{H}_{n_1,n_2}\hat{H}_{n_2,n_3}\hat{H}_{n3,0}}{n_1n_2n_3\omega^3}
\notag \\
&\quad\quad +\sum_{n\neq0} \frac{\hat{H}_{0,n}\hat{H}_{n,0}\hat{H}_{0,0}\hat{H}_{0,0}}{n^3\omega^3}
\notag \\
&\quad\quad -\sum_{n_1,n_2\neq0} \frac{\hat{H}_{0,n_1}\hat{H}_{n_1,0}\hat{H}_{0,n_2}\hat{H}_{n_2,0}}{n_1^2n_2\omega^3} 
\notag \\
&\quad\quad - \sum_{n_1,n_2\neq0} \frac{\hat{H}_{0,n_1}\hat{H}_{n_1,n_2}\hat{H}_{n_2,0}\hat{H}_{0,0}}{n_1n_2\omega^3} \left( \frac{1}{n_1}+\frac{1}{n_2} \right). 
\label{eq:BW3} 
\end{align}
%%%%%%%%%%%%%%%%%%%%%%%%%%%%%%%%%%%%%%%
Note that all the $n$-th order terms are the order of $(1/\omega)^n$. We neglect emergent many-body terms in $\hat{H}_{\rm BW}^{(2)}$ and $\hat{H}_{\rm BW}^{(3)}$ which have no influence on the Floquet single-particle states in the noninteracting systems~\cite{Mikami2016}.

\subsection{Cancellation of the first-order terms}
%%%%%%%%%%%%%%%%%%%%%%%%%%%%%%%%%%%%%%%
\begin{figure}[tbh]
\centering
\includegraphics[scale=1.0]{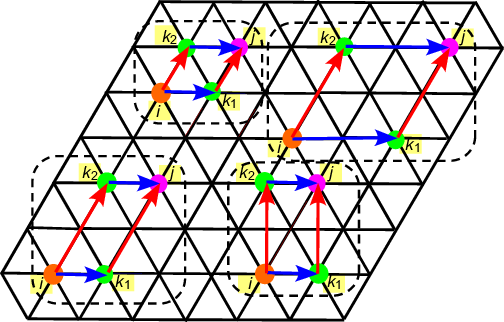}
\caption{Examples of pairs of three-site hopping paths $(i\rightarrow k_1\rightarrow j)$ and $(i\rightarrow k_2\rightarrow j)$ whose contributions to indirect hoppings from site $i$ to site $j$ cancel each other.}
 \label{Fig02}
\end{figure}
%%%%%%%%%%%%%%%%%%%%%%%%%%%%%%%%%%%%%%%
We discuss how the first-order terms of the Brillouin-Wigner expansion in the non-multipartite lattice electron systems vanish through mutual cancellation. The numerator of Eq.~(\ref{eq:BW1}) can be calculated by substituting Eq.~(\ref{eq:FCHm}) into Eq.~(\ref{eq:BW1}). The explicit calculations are done for $n>0$ as,
%%%%%%%%%%%%%%%%%%%%%%%%%%%%%%%%%%%%%%%
\begin{widetext}
\begin{align}
\left[ \hat{H}_n,\hat{H}_{-n} \right]
&=\sum_{i,j}\sum_{k,l} t_{ij}t_{kl}
\mathcal{J}_{-n}(\mathcal{A}_{ij}) \mathcal{J}_{n}(\mathcal{A}_{kl}) e^{-in\left(\theta_{ij}-\theta_{kl}\right)}
\left[ \hat{c}^\dagger_i\hat{c}_j,\hat{c}^\dagger_k\hat{c}_l \right] 
\notag\\
&=2i(-1)^{n+1} \sum_{i,j}\sum_k t_{ik}t_{kj}\mathcal{J}_{n}(\mathcal{A}_{ik})
\mathcal{J}_{n}(\mathcal{A}_{kj})
\sin\left\{ n\left( \theta_{ik}-\theta_{kj} \right) \right\} \hat{c}^\dagger_i\hat{c}_j.
\label{eq:BW1cancel}
\end{align}
\end{widetext}
%%%%%%%%%%%%%%%%%%%%%%%%%%%%%%%%%%%%%%%
Here we omit the spin indices for simplicity, but the generality of the argument is not compromised by consideration of the spin degrees of freedom. These terms describe three-site hoppings of conduction electrons from site $i$ to site $j$ via site $k$ ($i\rightarrow k \rightarrow j$) mediated by two nearest neighbor hoppings $t$. Importantly, in the two and three dimensional lattices without sublattice degrees of freedom, every three-site hopping path ($i\rightarrow k_1 \rightarrow j$) has its counterpart ($i\rightarrow k_2 \rightarrow j$), which satisfies the following relations,
%%%%%%%%%%%%%%%%%%%%%%%%%%%%%%%%%%%%%%%
\begin{align}
&t_{ik_1}t_{k_1j} \mathcal{J}_n(\mathcal{A}_{ik_1})\mathcal{J}_n(\mathcal{A}_{k_1j})
 =t_{ik_2}t_{k_2j} \mathcal{J}_n(\mathcal{A}_{ik_2})\mathcal{J}_n(\mathcal{A}_{k_2j}),
\\
&\theta_{ik_1}-\theta_{k_1j} = -(\theta_{ik_2}-\theta_{k_2j}).
\end{align}
%%%%%%%%%%%%%%%%%%%%%%%%%%%%%%%%%%%%%%%
These relations indicates that the summation over $k$ in the rightmost side of Eq.~(\ref{eq:BW1cancel}) leads to a perfect cancellation of the first-order terms. Examples of the pairs of three-site hopping paths $(i\rightarrow k_1\rightarrow j)$ and $(i\rightarrow k_2\rightarrow j)$ on the triangular lattice are shown in Fig.~\ref{Fig02}, whose contributions to indirect hoppings from site $i$ to site $j$ cancel each other. In contrast, the contributions from the first-order terms survive in the multipartite lattices, such as honeycomb lattices, Lieb lattices, and Kagome lattices. This situation is hardly changed even if further neighbor transfer integrals are considered for these lattices.

\subsection{Real-time simulations based on the time-dependent Schr\"{o}dinger equation}
In addition to the Floquet analysis, we also perform numerical simulations for real-time electron dynamics under photoirradiation based on the time-dependent Schr\"{o}dinger equation. The time-dependent Schr\"{o}dinger equation in Eq.~(\ref{eq:Schrodinger}) can be formally solved in the form,
\begin{widetext}
%%%%%%%%%%%%%%%%%%%%%%%%%%%%%%%%%%%%%%%
\begin{equation}
\ket{\psi_\nu({\bm k},\tau+\Delta\tau)} = \mathcal{T}\exp\left[ -i\int_\tau^{\tau+\Delta \tau} d\tau' \ \mathcal{H}_{\bm k}(\tau') \right] \ket{\psi_\nu({\bm k},\tau)},
\label{eq:Schodinger_solution}
\end{equation}
%%%%%%%%%%%%%%%%%%%%%%%%%%%%%%%%%%%%%%%
where $\ket{\psi_\nu({\bm k},\tau)}$ is the single-particle state of $\nu$th band with momentum $\bm k$ and time $\tau$, $\mathcal{H}_{\bm k}(\tau')$ denotes the ${\bm k}$-resolved Hamiltonian with time $\tau'$, and $\mathcal{T}$ is the time-ordering operator. The single-particle excitation spectrum is given by~\cite{Freericks2009,Sentef2015}
%%%%%%%%%%%%%%%%%%%%%%%%%%%%%%%%%%%%%%%
\begin{align}
A({\bm k},\varepsilon)={\rm Im} \sum_{i,\sigma} \int d\tau_1 d\tau_2 \;
s_{\rm pr}(\tau_1;\tau_{\rm pr},\sigma_{\rm pr}) \;s_{\rm pr}(\tau_2;\tau_{\rm pr},\sigma_{\rm pr})
\; e^{i\varepsilon(\tau_1-\tau_2)}
\left[ G^<_{{\bm k},ii\sigma\sigma}(\tau_1,\tau_2) - G^>_{{\bm k},ii\sigma\sigma}(\tau_1,\tau_2) \right],
\label{eq:Aqwformula}
\end{align}
%%%%%%%%%%%%%%%%%%%%%%%%%%%%%%%%%%%%%%%
\end{widetext}
with
%%%%%%%%%%%%%%%%%%%%%%%%%%%%%%%%%%%%%%%
\begin{align}
G^<_{{\bm k},ij\sigma\sigma'}(\tau_1,\tau_2) &= i\braket{\hat{c}^\dagger_{{\bm k},j\sigma'}(\tau_2) \hat{c}_{{\bm k},i\sigma}(\tau_1)}, 
\\
G^>_{{\bm k},ij\sigma\sigma'}(\tau_1,\tau_2) &= -i\braket{\hat{c}_{{\bm k},i\sigma}(\tau_1) \hat{c}^\dagger_{{\bm k},j\sigma'}(\tau_2)}.
\end{align}
%%%%%%%%%%%%%%%%%%%%%%%%%%%%%%%%%%%%%%%
The lesser and greater Green's functions $G^<_{{\bm k},ii\sigma\sigma}(\tau_1,\tau_2)$ and $G^>_{{\bm k},ii\sigma\sigma}(\tau_1,\tau_2)$ correspond to occupied and unoccupied states of the $i$th site with momentum $\bm k$ and spin $\sigma(=\uparrow,\downarrow)$, respectively. $s_{\rm pr}(t;\tau_{\rm pr},\sigma_{\rm pr})$ denotes the normalized Gaussian wavepacket of the probe pulse given in the form,
%%%%%%%%%%%%%%%%%%%%%%%%%%%%%%%%%%%%%%%
\begin{equation}
s_{\rm pr}(\tau;\tau_{\rm pr},\sigma_{\rm pr}) = \frac{1}{\sqrt{2\pi}\sigma_{\rm pr}} \exp\left[ -\frac{(\tau-\tau_{\rm pr})^2}{2\sigma_{\rm pr}^2} \right],
\end{equation}
%%%%%%%%%%%%%%%%%%%%%%%%%%%%%%%%%%%%%%%
where $\tau_{\rm pr}$ and $\sigma_{\rm pr}$ are the pulse center and the pulse width, respectively. The effect of pump pulse, which is treated as a time-periodic external field in the framework of the Floquet theory, is also taken into account by means of the Peierls substitution $\bm k \rightarrow \bm k + \bm A(\tau)$. The vector potential that mimics a circularly polarized pump pulse with the Gaussian envelope is given by,
%%%%%%%%%%%%%%%%%%%%%%%%%%%%%%%%%%%%%%%
\begin{equation}
\bm A_{\rm pu}(\tau)=\frac{E_\omega}{\omega} \exp\left[ -\frac{(\tau-\tau_{\rm pu})^2}{2\sigma_{\rm pu}^2} \right] \left( {\bm e}_1\cos\omega\tau + {\bm e}_2\sin\omega\tau \right),
\label{eq:PumpLaser}
\end{equation}
%%%%%%%%%%%%%%%%%%%%%%%%%%%%%%%%%%%%%%%
with maximum electric field strength of $E_\omega$ and the frequency of $\omega$. Here $\tau_{\rm pu}$ and $\sigma_{\rm pu}$ are the pulse center and the pulse width, respectively. We set $(\tau_{\rm pu},\tau_{\rm pr},\sigma_{\rm pu},\sigma_{\rm pr})=(500T,500T,76T,25T)$ in this work. We solve the time-dependent Sch\"{o}dinger equation in Eq.(\ref{eq:Schodinger_solution}) in the time window of $t\in[0,1000T]$ with the time step of $\Delta\tau=T/800$ where $T=2\pi/\omega$. The initial states $\left\{ \ket{\psi_\nu({\bm k},0)} \right\}$ are set to be the eigenstates of $\mathcal{H}_{\rm IKLM}({\bm k})$ in equilibrium where $\bm A(\tau)=0$. The exponential function in Eq.~(\ref{eq:Schodinger_solution}) is expanded up to the 20th order.
%The proof of a positive definiteness of the spectral function $A(\bm k, \varepsilon)$ in the Floquet theory is given in Ref. 

\section{Results}
\subsection{Phase diagram}
%%%%%%%%%%%%%%%%%%%%%%%%%%%%%%%%%%%%%%%
\begin{figure*}[tbh]
\centering
\includegraphics[scale=1.0]{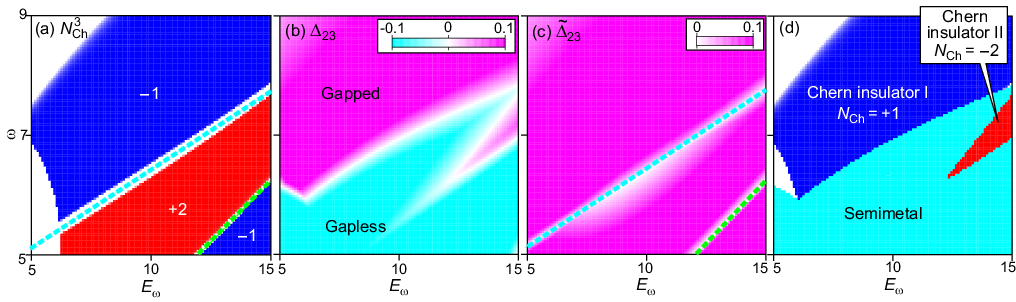}
\caption{(a)-(c) $E_\omega$-$\omega$ profiles of (a) the Chern number of the third band $N_{\rm Ch}^3$, (b) the indirect band gap $\Delta_{23}$, and (c) the direct band gap $\tilde{\Delta}_{23}$ calculated by diagonalizing the truncated Floquet Hamiltonian $\mathcal{H}_{\rm F}(\bm k)-\omega\mathcal{M}_{\rm m}$. (d) Nonequilibrium phase diagram in the plane of light amplitude $E_\omega$ and frequency $\omega$ under irradiation with circularly polarized light, which is constructed from the $E_\omega$-$\omega$ profiles of the physical quantities in (a)-(c).}
\label{Fig03}
\end{figure*}
%%%%%%%%%%%%%%%%%%%%%%%%%%%%%%%%%%%%%%%
We first study nonequilibrium steady states of electrons under irradiation with circularly polarized light by using the truncated Floquet Hamiltonian $\mathcal{H}_{\rm F}(\bm k)-\omega\mathcal{M}_{\rm m}$. To construct a phase diagram, we calculate $E_\omega$-$\omega$ profiles of several physical quantities, i.e., Chern number of the third band $N_{\rm Ch}^3$, indirect band gap $\Delta_{23}$, and direct band gap $\tilde{\Delta}_{23}$ by diagonalizing the truncated Floquet Hamiltonian. The obtained profiles are shown in Figs.~\ref{Fig03}(b)-(d), respectively. Note that we adopt the natural units $e=\hbar=c=1$ for the following calculations. Table~\ref{Table01} summarizes the unit conversions when we assume $t$=0.1 eV and $a$=5 \AA for the transfer integral and the lattice constant, respectively.
%%%%%%%%%%%%%%%%%%%%%%%%%%%%%%%%%%%%%%%
\begin{table*}[tbh]
\centering
\caption{Unit conversions when the transfer integral is $t$=0.1 eV and the lattice constant is $a$=5 \AA. The variables with (without) tilde denote dimensionful (dimensionless) quantities.}
\begin{tabular}{ccc} \hline
Quantity & Dimensionless quantity & Corresponding values \\ \hline
Frequency & $\omega=\hbar\tilde{\omega}/t$=1 & $\tilde{\omega}$=24.2 THz \\
Light electric field & $E_\omega=ea\tilde{E}_\omega/t$=1 & $\tilde{E}_\omega$=2 MV cm$^{-1}$ \\
Time & $\tau=\tilde{\tau}t/\hbar$=1 & $\tilde{\tau}$=6.6 fs \\
Temperature & $T_{\rm hr}=k_{\rm B}\tilde{T}_{\rm hr}/t$=1 & $\tilde{T}_{\rm hr}$=1163 K \\
\hline
\end{tabular}
\label{Table01}
\end{table*}
%%%%%%%%%%%%%%%%%%%%%%%%%%%%%%%%%%%%%%%

The original six bands are divided into two band sets with three bands each separated by the exchange gap. Importantly, sums of the Chern numbers $N_{\rm Ch}^\nu$ over the three bands within a band set become zero, that is, $\sum_{\nu=1}^3 N_{\rm Ch}^\nu=0$ and $\sum_{\nu=4}^6 N_{\rm Ch}^\nu=0$. Accordingly, when the electron filling is nearly 1/3 with only the lowest two of six bands originally filled with electrons, the Chern number of the system $N_{\rm Ch}=\sum_{\nu \in {\rm filled}}N_{\rm Ch}^\nu$ is related with $N_{\rm Ch}^3$ as $N_{\rm Ch}=-N_{\rm Ch}^3$. In Fig.~\ref{Fig03}(a), we find two regions with $N_{\rm Ch}^3=-1$ ($N_{\rm Ch}=+1$) separated by a region with $N_{\rm Ch}^3=+2$ ($N_{\rm Ch}=-2$). 

The $N_{\rm Ch}^3=-1$ phase in the high-frequency regime is caused by a gap opening at the K point as discussed in Sec.~IV.B. As the system approaches from the high-frequency $N_{\rm Ch}^3=-1$ phase to the $N_{\rm Ch}^3=+2$ phase, the band gap at the M point gradually closes and completely vanishes at the boundary between these two phases. Subsequently, the gap reopens at the M point as the system enters the $N_{\rm Ch}^3=+2$ phase as described in Sec.~IV.C. On the contrary, the boundary between the $N_{\rm Ch}^3=+2$ phase and the low-frequency $N_{\rm Ch}^3=-1$ phase corresponds to the point at which the flat band appears with vanishing bandwidth due to the dynamical localization. According to the Floquet theory, the transfer integrals under a periodic drive are renormalized by a factor of the Bessel function $\mathcal{J}_0(E_\omega/\omega)$, and the factor becomes zero at this phase boundary. We also note that the Chern number is ill-defined in the white regions because the band gap is too small.

The $E_\omega$-$\omega$ profiles of two types of band gaps (indirect and direct gaps) between the second and the third bands are shown in Figs.~\ref{Fig03}(b) and \ref{Fig03}(c). The indirect and direct band gaps $\Delta_{\mu\nu}$ and $\tilde{\Delta}_{\mu\nu}$ between the $\mu$th and $\nu$th bands are defined by,
%%%%%%%%%%%%%%%%%%%%%%%%%%%%%%%%%%%%%%%
\begin{align}
&\Delta_{\mu\nu} = \min_{{\bm k}\in{\rm BZ}}\left[ \varepsilon_\nu^0({\bm k}) \right] 
- \max_{{\bm k}\in{\rm BZ}}\left[ \varepsilon_\mu^0({\bm k}) \right], 
\\
&\tilde{\Delta}_{\mu\nu} = \min_{{\bm k}\in{\rm BZ}}
\left[ \varepsilon_\nu^0({\bm k}) - \varepsilon_\mu^0({\bm k}) \right].
\end{align}
%%%%%%%%%%%%%%%%%%%%%%%%%%%%%%%%%%%%%%%
According to $\Delta_{23}$, we judge whether the system is gapped/insulating ($\Delta_{23}>0$) or gapless/metallic ($\Delta_{23}<0$). On the other hand, we capture the closing of direct band gap according to $\tilde{\Delta}_{23}$, which becomes zero on the dashed lines in Fig.~\ref{Fig03}(c). These two lines correspond to the two phase boundaries in Fig.~\ref{Fig03}(a), i.e., a phase boundary at which the gap closes at M points and another phase boundary at which the doubly degenerate up-spin-polarized bands appear.

The $E_\omega$-$\omega$ profiles of the physical quantities in Figs.~\ref{Fig03}(a)-(c) are summarized into a phase diagram in Fig.~\ref{Fig03}(d). We assign the region with $\Delta_{23}<0$ to a semimetal phase irrespective of the value of $N_{\rm Ch}$. On the contrary, we assign the regions with $\Delta_{23}>0$ and nonzero quantized $N_{\rm Ch}$ to a Chern insulator phase. We have two types of Chern insulator phases, i.e., the phase I with $N_{\rm Ch}=+1$ and the phase II with $N_{\rm Ch}=-2$. %%As will be discussed l in Sec.~IV.E, it turns out that the sign of $\Delta_{23}$ or whether the system is semimetallic or insulating does not affect the quantization of Hall conductivity so much. Therefore, less attention will paid to the sign of $\Delta_{23}$ in the following discussion.

\subsection{Single-particle spectra}
%%%%%%%%%%%%%%%%%%%%%%%%%%%%%%%%%%%%%%%
\begin{figure*}[tbh]
\centering
\includegraphics[scale=1.0]{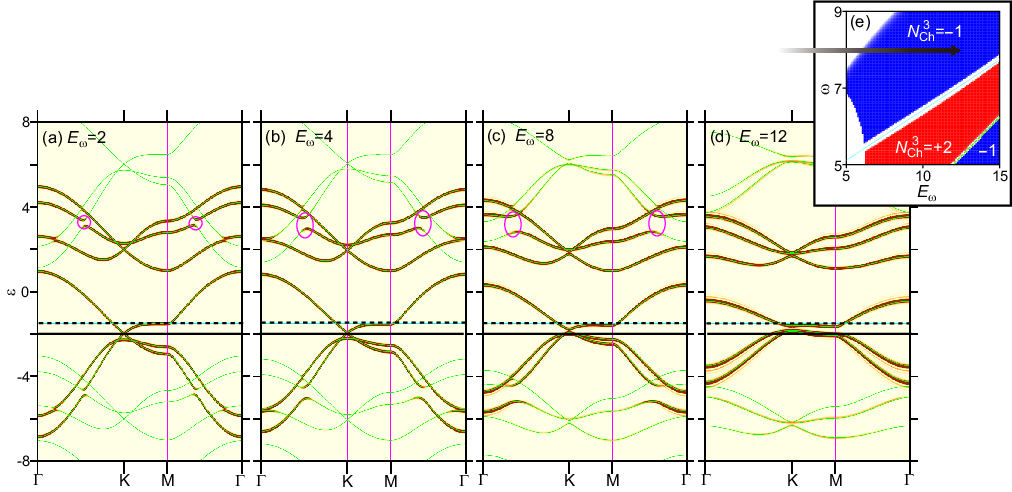}
\caption{(a)-(d) Single-particle spectra $A(\bm q,\varepsilon)$ for various values of $E_\omega$, i.e., (a) $E_\omega$=2, (b) $E_\omega$=4, (c) $E_\omega$=8, (d) $E_\omega$=12, respectively, when $\omega$=8. The horizontal solid lines represent the energy level of Dirac points at K point in equilibrium, while the horizontal dashed lines represent the chemical potential ($\mu=-1.5$) used in the present simulations. (e) Variation of $E_\omega$ from (a) to (d) is indicated by a horizontal thick arrow.}
\label{Fig04}
\end{figure*}
%%%%%%%%%%%%%%%%%%%%%%%%%%%%%%%%%%%%%%%
%%%%%%%%%%%%%%%%%%%%%%%%%%%%%%%%%%%%%%%
\begin{figure*}[tbh]
\centering
\includegraphics[scale=1.0]{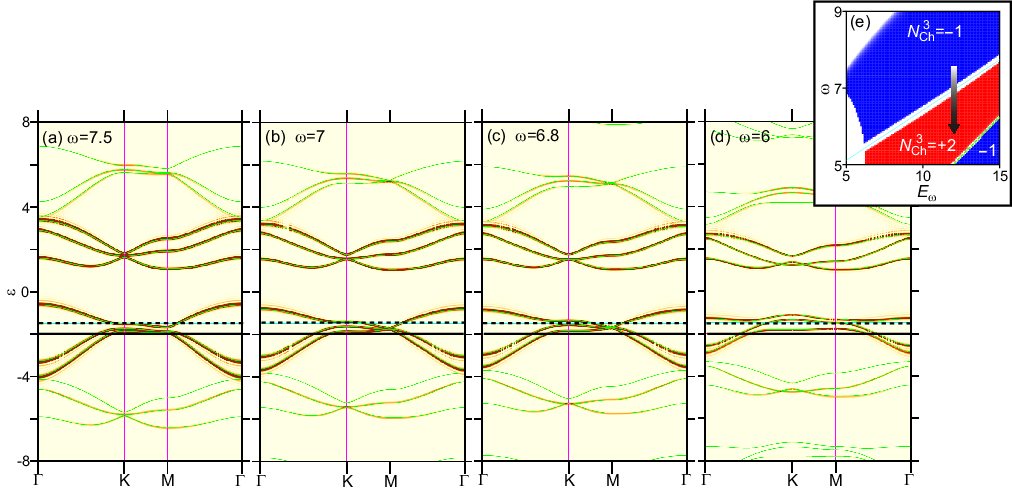}
\caption{(a)-(d) Single-particle spectra $A(\bm q,\varepsilon)$ for various values of $\omega$, i.e., (a) $\omega$=7.5 (b) $\omega$=7, (c) $\omega$=6.8, (d) $\omega$=6, respectively, when $E_\omega$=12. The horizontal solid lines represent the energy level of Dirac points at K point in equilibrium, while the horizontal dashed lines represent the chemical potential ($\mu=-1.5$) used in the present simulations. (e) Variation of $\omega$ from (a) to (d) is indicated by a vertical thick arrow.}
\label{Fig05}
\end{figure*}
%%%%%%%%%%%%%%%%%%%%%%%%%%%%%%%%%%%%%%%
To understand how these Floquet Chern insulator phases appear, we investigate the Floquet band structures under photoirradiation. Figures~\ref{Fig04}(a)-(d) show the single-particle spectra $A(\bm q,\varepsilon)$ calculated by real-time simulations of the time-dependent Schr\"{o}dinger equation (color) and the Floquet band structures calculated by diagonalization of the truncated Floquet Hamiltonian $\mathcal{H}_{\rm F}(\bm k)-\omega\mathcal{M}_{\rm m}$ (green solid lines) for various values of $E_\omega$ when $\omega$ is fixed at 8. The value of $E_\omega$ is increased as indicated by a horizontal arrow in Fig.~\ref{Fig04}(e). 

Figure~\ref{Fig04}(a) indicates that the Dirac point located at K point is almost gapless when the light amplitude is as small as $E_\omega$=2. The Dirac gap is gradually opened with increasing $E_\omega$, and eventually the Dirac point is apparently gapped when the light amplitude is as large as $E_\omega$=12 as seen in Fig.~\ref{Fig04}(d). In the presence of the gap, the two lowest bands acquire a nonzero Chern number of $N_{\rm Ch}$=1 in total. Note that the band gap between the second and third bands on the M point is always opened irrespective of the value of $E_\omega$. 

In the Floquet band structures, the band set, which almost perfectly overlaps the single-particle spectrum, corresponds to the electron states with zero-photon absorption. In addition to this original band set, the equivalent band sets repeatedly appear with an energy interval of $\omega$, which correspond to the $n$-photon absorbed (emitted) electron states and are referred to as the Floquet sidebands. When the light amplitude is small as $E_\omega$=2 in Fig.~\ref{Fig04}(a), we find that the original band set and the Floquet sideband sets located right above and below it partially overlap, because the light frequency of $\omega=8$ is smaller than the bandwidth of $W \sim 12$ (i.e., $\omega<W$). This situation is referred to as the on-resonant case. 

The overlap of band sets occurs also when $E_\omega$=4 and $E_\omega$=8. In this on-resonant case, band-anticrossing occurs at specific points indicated by solid circles. Interestingly, the spectral weight $A(\bm q,\varepsilon)$ is extended to the sidebands at the band-anticrossing point, which indicates that the Floquet sidebands are partially occupied by electrons. On the contrary, there are some points at which band crossing instead of anticrossing occurs. At these band-crossing points, the spectral weight $A(\bm q,\varepsilon)$ does not exhibit particular change or anomaly. The crossing and anticrossing of the bands are governed by the structure of the Hamiltonian matrix. The Hamiltonian matrix $\mathcal{H}_{\rm IKLM}$ can be block-diagonalized into independent up-spin and down-spin blocks as $\mathcal{H}_{\rm IKLM}=\mathcal{H}_\uparrow\oplus\mathcal{H}_\downarrow$. When two bands belong to the same (different) spin blocks, the anticrossing (crossing) occurs when they cross.

Next we discuss the variation of band structures upon another topological phase transition, which occurs from the $N_{\rm Ch}=+1$ phase to the $N_{\rm Ch}=-2$ phase with decreasing $\omega$ when $E_\omega$ is fixed at 12. Figure~\ref{Fig05}(a) indicates that both K and M points are gapped when $\omega$=7.5, and the system has a Chern number of $N_{\rm Ch}=+1$ associated with the photoinduced gapped Dirac point at the K point. As $\omega$ decreases, the band gap at the M point gradually decreases. Indeed, as seen in Figs.~\ref{Fig05}(b) and (c), the band gap at the M point is almost closed when $\omega=7$ and $\omega=6.8$. With further decreasing $\omega$, the band gap start opening again after the system enters the $N_{\rm Ch}=-2$ phase. As seen in Fig.~\ref{Fig05}(d), a clear gap opens up when $\omega=6$. 

The Chern number of $N_{\rm Ch}=-2$ in the system after the gap reopening at the M point can be understood as follows. First, the Dirac-gap opening at the K point gives the Chern number of $+1$ to the lowest two bands, i.e., $N_{\rm Ch}^1+N_{\rm Ch}^2=+1$. Subsequently, the closing and reopening of gap at the M points give additional Chern number of $-1$ at each M point. Because the system has three independent M points in the hexagonal Brillouin zone of the up-up-down ferrimagnetic order, the total Chern number becomes $N_{\rm Ch}=(+1)+(-1)\times3=-2$.

\subsection{Hall conductivity}
%%%%%%%%%%%%%%%%%%%%%%%%%%%%%%%%%%%%%%%
\begin{figure}[tbh]
\centering
\includegraphics[scale=1.0]{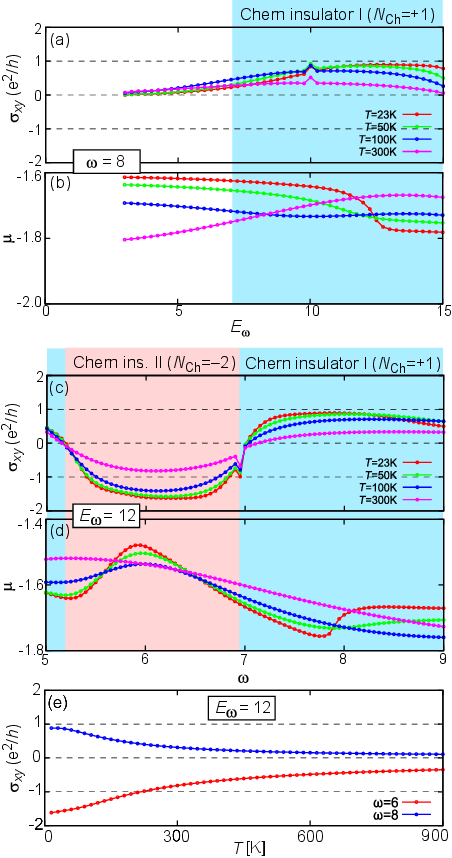}
\caption{(a) Hall conductivity $\sigma_{xy}$ and (b) Chemical potential $\mu$ for the electron filling of $n_{\rm e}$=0.34 in the photodriven system as functions of $E_\omega$ when $\omega$=8. (c),~(d) Those as functions of $\omega$ when $E_\omega$=12. (e) $\sigma_{xy}$ as a function of temperature when $E_\omega=12$ and $\omega=6,8$.}
\label{Fig06}
\end{figure}
%%%%%%%%%%%%%%%%%%%%%%%%%%%%%%%%%%%%%%%
It is predicted that these two Floquet Chern insulator phases in the photodriven ferrimagnetic system can be detected and distinguished experimentally by the measurement of Hall conductivity $\sigma_{xy}$. Figure~\ref{Fig06}(a) shows the $E_\omega$-dependence of $\sigma_{xy}$ calculated using the Keldysh Green's-function formalism when $\omega$ is fixed at 8. Here, Fig.~\ref{Fig06}(b) shows the calculated chemical potential $\mu$ when the electron filling is $n_{\rm e}$=0.34. In Fig.~\ref{Fig06}(a), we find that $\sigma_{xy}$ is almost zero when $E_\omega$ is small as $E_\omega \lessapprox 4$. As $E_\omega$ increases, $\sigma_{xy}$ increases gradually. When $E_\omega\sim 12$ inside the $N_{\rm Ch}=1$ phase, $\sigma_{xy}$ is nearly quantized to $e^2/h$ at a low temperature of $T$=23 K, which corresponds to the Chern number of $N_{\rm Ch}=+1$ in the Floquet Chern insulator phase I. Such a quantization of $\sigma_{xy}$ cannot be seen at a higher temperature of $T$=300 K. However, we still observe a positive nonzero $\sigma_{xy}$ as large as 20\% of the quantized value at $T$=300 K. 

On the other hand, Figs.~\ref{Fig06}(c) and (d) show the $\omega$-dependence of $\sigma_{xy}$ and $\mu$ when $E_\omega$ is fixed at 12. We find that nearly quantized values of $\sigma_{xy}$ of $-2e^2/h$ and $e^2/h$ are observed in the two Floquet Chern insulator phases with $N_{\rm Ch}=-2$ and $N_{\rm Ch}=+1$, respectively, at low temperatures ($T$=23 K), whereas the quantization is obscure at higher temperatures ($T$=300 K). A clear sign change of $\sigma_{xy}$ from negative to positive is observed when the system enters from the $N_{\rm Ch}=-2$ phase to the $N_{\rm Ch}=+1$ phase with increasing $\omega$. Interestingly, this sign change is observed not only at $T$=23K but also at $T$=300 K. Moreover, Fig.~\ref{Fig06}(e) indicates that the sign change survives even at 900 K. The heating is unavoidable in real experiments with light irradiation, but this result indicates that the sign change is robust against rise in temperature, which is a favorable property for the experimental observation.

In Figs.~\ref{Fig06}(c), we can find that the quantization of the Hall currents are more pronounced in the $N_{\rm Ch}=+1$ regions compared to the $N_{\rm Ch}=-2$ regions at low temperatures. One possible scenario to describe the reason of the difference in quantization is the difference in nonequilibrium band structures between the $N_{\rm Ch}=+1$ phase and the $N_{\rm Ch}=-2$ phase. In fact, Figs.~\ref{Fig03}(a) and \ref{Fig03}(b) indicate that the indirect bandgap $\Delta_{23}$ is positive for most of the $N_{\rm Ch}=+1$ regions while it is negative for most of the $N_{\rm Ch}=-2$ regions. Considering that the negative indirect bandgap makes the quantized Hall signatures obscure, it makes sense that the quantization of the Hall currents in $N_{\rm Ch}=-2$ is less pronounced than those in $N_{\rm Ch}=+1$.

\subsection{Analyses based on the high-frequency expansion}
%%%%%%%%%%%%%%%%%%%%%%%%%%%%%%%%%%%%%%%
\begin{figure}[tbh]
\centering
\includegraphics[scale=1.0]{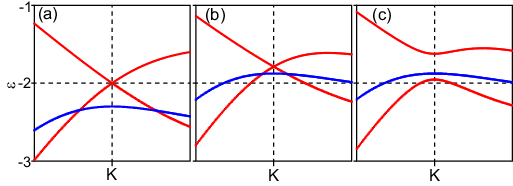}
\caption{Contributions from terms of the high-frequency expansion (the Brillouin-Wigner expansion) to the Floquet-band structure in the photoirradiated system. The band dispersion relations around the K point are calculated using the Brillouin-Wigner Hamiltonian $\hat{H}_{\rm BW}$ in Eq.~(\ref{eq:BWE}) up to (a) the first-, (b) the second-, and (c) the third-order terms with respect to $1/\omega$ when $E_\omega$=6 and $\omega$=6.}
\label{Fig07}
\end{figure}
%%%%%%%%%%%%%%%%%%%%%%%%%%%%%%%%%%%%%%%
As discussed in Sec.III.C, the Floquet engineering has often been performed using the effective Floquet Hamiltonian obtained by the Brillouin-Wigner expansion in the high-frequency limit. More specifically, Hamiltonians composed of up to the first-order terms of the expansion with respect to $1/\omega$ have been frequently used for the research. However, it is known that the first-order terms usually vanish because of the cancellation of equivalent paths having phases with opposite signs. In multipartite lattices, we can avoid this cancellation to obtain a nonzero contribution from the first-order terms. On the contrary, the cancellation cannot be avoided in simple lattices such as square and triangular lattices even in the presence of extrinsic sublattice degrees of freedom introduced by long-range orders of spins and/or charges. Thereby, Floquet topological electron states and photoinduced topological phase transition cannot be expected in electron systems on simple lattices within the crude approximation based on the high-frequency expansion up to the first-order. This is a reason why several multipartite lattices such as Kagome lattices, honeycomb lattices, and Lieb lattices have been studied in the research of Floquet engineering, whereas simple square lattices and triangular lattices have not been in a scope of the research.

However, as discussed above, we have obtained rich Floquet Chern insulator phases in the triangular Kondo-lattice model by the analyses based on direct diagonalization of the truncated effective Floquet Hamiltonian. In the present system, a ferrimagnetic order has been assumed to realize the Dirac-cone bands around the K point. The above-argued cancellation of the first-order terms also occurs even in the presence of sublattice degrees of freedom due to the ferrimagnetic spin order. In fact, the Floquet Chern insulator phases in the present triangular-lattice system are attributable to the higher-order terms of the expansion. 
%%\textcolor{red}{Note that, although the high-frequency expansion does not have enough quantitative accuracy for the frequency $\omega=6$ or 8 which are smaller than the bandwidth $W\sim12$ and are frequently used in this paper, it is still helpful to qualitatively understand the contribution of photoirradiation to the band structure modulation by relating it with lattice geometry.} 

To demonstrate the substantial role of the higher-order terms, we calculate the band structure around the K point under irradiation with circularly polarized light when $E_\omega=6$ and $\omega=6$. Figure~\ref{Fig07}(a) shows the band structure calculated using a Hamiltonian matrix composed of up to the first-order terms only, that is, $\hat{H}_{\rm BW}=\hat{H}_{\rm BW}^{(0)}+\hat{H}_{\rm BW}^{(1)}$. This band structure is nearly the same with that in equilibrium without photoirradiation in Fig.~\ref{Fig01}(c), because the zeroth-order term $\hat{H}_{\rm BW}^{(0)}$ corresponds to the time-averaged Hamiltonian and the first-order term $\hat{H}_{\rm BW}^{(1)}$ vanishes due to the perfect cancellation.

In Fig.~\ref{Fig07}(b), we show the band structure calculated using a Hamiltonian matrix including up to the second-order terms, that is, $\hat{H}_{\rm BW}=\sum_{n=0}^2 \hat{H}_{\rm BW}^{(n)}$. The overall band structure shifts upward along the energy axis, indicating that the second-order term $\hat{H}_{\rm BW}^{(2)}$ has a nonzero contribution, but it cannot open a gap at the Dirac point. The band structure calculated for $\hat{H}_{\rm BW}=\sum_{n=0}^3 \hat{H}_{\rm BW}^{(n)}$ in Fig.~\ref{Fig07}(c), in contrast, shows a gapped Dirac point. This indicates that the third-order term $\hat{H}_{\rm BW}^{(3)}$ is the lowest-order term required to open a gap at the Dirac point. Owing to this gap opening, the system attains a nonzero Chern number of $N_{\rm Ch}=+1$.

These results clearly demonstrate that the higher-order terms of the Brillouin-Wigner expansion instead of the usually considered first-order terms are relevant to the gap opening at the Dirac point, which results in the photoinduced topological phase transition and rich Floquet topological phases. This aspect is expected to widen the target materials of the Floquet engineering and to enhance the possibility of the research.

\section{Summary}

\textcolor{red}{As we discussed earlier in this paper, the main targets of the Floquet engineering have been the multipartite lattices, e.g., honeycomb, Kagome, and Lieb lattices in which the sublattice degrees of freedom are coming only from lattice geometry, in other words, originally imprinted even without considering spin or orbital degrees of freedom of electrons. We here have demonstrated that such ``lattice geometry-induced" or ``originally imprinted” sublattice degrees of freedom are not necessarily required for the Floquet engineering and it could be done even in the monopartite systems, as long as they have “non-imprinted” sublattice degrees of freedom induced by the spontaneous orderings of spins (or even orbitals). In the latter case, not the lowest but higher-order terms in the high-frequency expansion give us the feasibility of the Floquet engineering.}

More specifically, as a typical example of the latter case, we have theoretically studied the effects of photoirradiation with circularly polarized light on the Dirac half-metal state in the triangular Kondo-lattice model with a three-sublattice ferrimagnetic order. By applying the Floquet analysis based on the truncated Floquet Hamiltonian, we have found that two types of Floquet Chern insulator phases with distinct Chern numbers of $N_{\rm Ch}=+1$ and $N_{\rm Ch}=-2$ appear as nonequilibrium steady states, which originate from the band gap formation/closing at distinct momentum points. By calculating the Hall conductivity in the photodriven system using the Keldysh Green's function formalism, we have revealed that these two Floquet Chern insulator phases can be experimentally detected and are distinguishable by measurements of the Hall conductivity. Specifically, it has been revealed that the Hall conductivity takes nearly quantized values of $e^2/h$ and $-2e^2/h$ with opposite signs in the respective phases. It has also been elucidated that these nonequilibrium topological phases come from the higher-order terms in the Brillouin-Wigner expansion in the high-frequency limit, which is in striking contrast to usually discussed Floquet Chern insulator phases originating from the lowest-order terms of the expansion. Because the first-order terms generally cancel out and vanish in simple non-multipartite lattices, research of the Floquet engineering has been performed by taking several multipartite lattices, e.g., the Kagome lattices, the honeycomb lattices, and the Lieb lattices. However, the present work has revealed that the higher-order terms, which have nonzero contributions even in the non-multipartite lattices, can induce the photoinduced topological phase transitions and the Floquet topological electron phases. This aspect indicates that various lattice electron models on simple lattices such as square lattices and triangular lattices can also be within a scope of the Floquet engineering. We expect that the predicted Floquet Chern insulator phases might be observed in triangular ferrimagnets $R$Fe$_2$O$_4$ ($R$=Yb, Lu, Er)~\cite{Tanaka1989,Iida1993,Kito1995a,Kito1995b} under irradiation with circularly polarized light. The present work will widen the list of candidate target materials/systems for the Floquet engineering and enhance the possibility of research in this field.

\section{Acknowledgment}
%R.E. is grateful to Atsushi Ono for sharing the methodology to determine the chemical potential in the Keldysh Green's function calculations by using the bisection method. 
R.E. is grateful to Atsushi Ono for useful discussion. This work was supported by CREST, the Japan Science and Technology Agency (Grant No. JPMJCR20T1), JSPS KAKENHI (Grants No. JP20H00337, No. JP23H04522, and No. 24H02231), and Waseda University Grant for Special Research Projects (2023C-140 and 2024C-153). R.E. was supported by a Grant-in-Aid for JSPS Fellows (Grant No. 23KJ2047). Numerical calculations were in part performed at the Supercomputer Center, Institute for Solid State Physics, University of Tokyo.

\end{document}